\documentclass[final,1p,times]{elsarticle} % do not change
\usepackage{graphicx} % do not change
\usepackage{amssymb} % do not change
\usepackage{amsthm} % do not change
\usepackage{lineno} % do not change

\journal{Nuclear Physics A} % do not change
\begin{document} % do not change

\begin{frontmatter} % do not change

%% QM09Author: please enter your  
%% Title, author and address info here; please do not use footnotes

% Your Title - please modify
\title{
Jet Quenching in Evolving QGP Medium
}

% Principle author, and co-authors - please modify
\author{ Sangyong Jeon$^a$ }

% Address - please modify
% note that if you have authors from several institutions, we recommend
% labelling these [a], [b], [c] etc.
\address[a]{
Dept.~of Physics, McGill University,
3600 rue Rutherford,
Montreal QC H3A02T8, Canada
}

\begin{abstract} % do not change
A short summary of different approaches to the parton energy loss problem is
given. A particular attention is paid to the differences between
various models. A possible solution to the problem of distinguishing
competing approaches is discussed.
\end{abstract} % do not change

\end{frontmatter} % do not change

%% QM09: we keep linenumbers at least for initial version
%\linenumbers % do not change

%% start of main text - please modify. Below is a sub-set (single section) 
%% of an earlier proceedings that show how one can handle references 
%% and figures etc.
%%\section{}\label{}

\section{Introduction}

For more than 2 decades, one of the most important goals in high-energy
nuclear physics has been the study of the quark-gluon plasma (QGP for short).
The existence of this deconfined phase of quarks and gluons has been
unequivocally shown in many Lattice QCD (Quantum Chromodynamics) studies 
(for a recent study see Ref.~\cite{Bazavov:2009zn}).
In those studies, the phase transition (cross-over) temperature between 
the hadronic phase and the QGP phase is shown to be roughly about 
200 MeV when 3 light species ($u, d, s$) of dynamic quarks are taken 
into account. 
This value of transition temperature is actually somewhat of a conundrum.
On the one hand, this temperature, although exceeding two trillion
Kelvin, is nonetheless low enough to be accessible in accelerator experiments.
On the other hand, the energy scale corresponding to this temperature 
is too low for QCD to be perturbative, making analytic
calculations difficult.

Nevertheless, combining the results from lattice QCD calculations, insights
from perturbative thermal QCD calculations and also thermodynamics, 
we do know much about qualitative features of this new phase of the matter. 
Accordingly, many researchers have proposed many different signals 
of the formation of the QGP in heavy ion collisions.
Many of these proposed signals utilize the fact that
the energy density of QGP is extremely high.
At $T=200\,{\rm MeV}$, the energy density easily exceeds
$2\,{\rm GeV}/{\rm fm}^3$. 
Systems created at the Relativistic Heavy Ion Collider (RHIC) 
can reach maximum temperature of about $500\,{\rm MeV}$. 
At this
temperature, the energy density of QGP (composed of gluons and $u, d, s$
quarks) can be as high as $100\,{\rm GeV}/{\rm fm}^3$.

Not surprisingly, the three most prominent QGP signals that emerged from 
RHIC experiments --
the strong elliptic flow, the quenching of the 
high-energy (jet-) particles and the emergence of the medium-generated
photons -- are all measures of the high energy density and corresponding high
pressure.
Among them, the jet-quenching phenomenon is perhaps the most direct
observation of the high energy density. Intuitively, it is easy to see that
it will take an extremely dense matter to stop a particle with 
an extremely high energy.
In this proceeding, a short summary of the various theoretical
concepts that go into calculating the jet quenching 
(equivalently, parton energy loss) is presented.
It is, of course, impossible to do justice to the vast amount of work 
performed by many different researchers in this short proceeding.
What I will do mainly is to use the McGill-AMY approach 
\cite{Arnold:2002ja,Jeon:2003gi,Turbide:2005fk,
Qin:2007rn,Gale:2008wf,Turbide:2007mi,Qin:2009bk,Schenke:2009ik}
that I am most
familiar with as an illustrative example and highlight the differences
between this and other approaches where possible.
Interested readers are directed to a more comprehensive review already in print
\cite{Bass:2008rv}
and let me apologize here to people whose interesting works
I cannot fully cover here for the lack of space.

\section{Schematic Idea of Jet Quenching}

\begin{figure}[ht]
\centering
\includegraphics[width=0.30\textwidth]{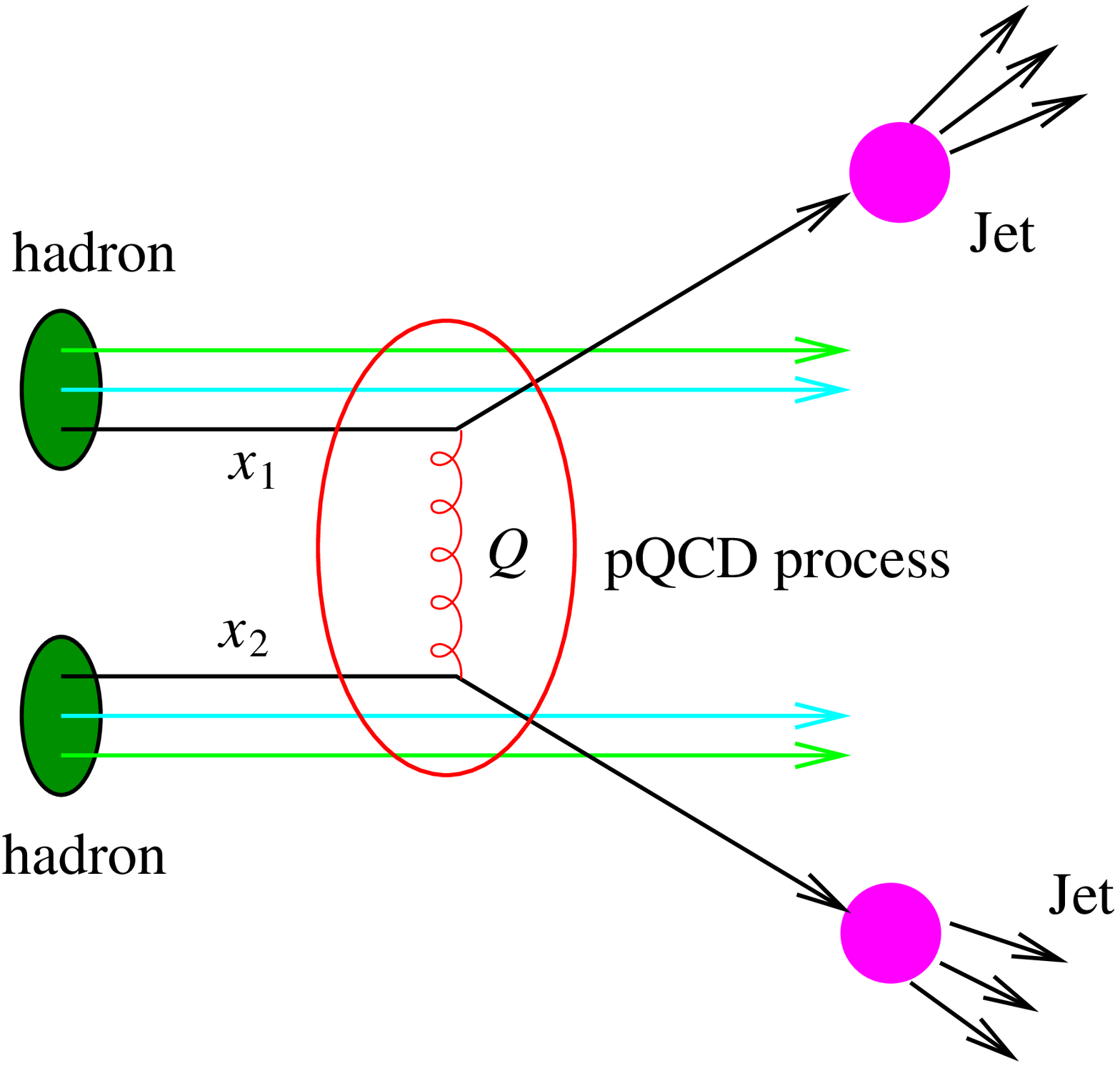}	       
\hspace{1.2cm}
\includegraphics[width=0.30\textwidth]{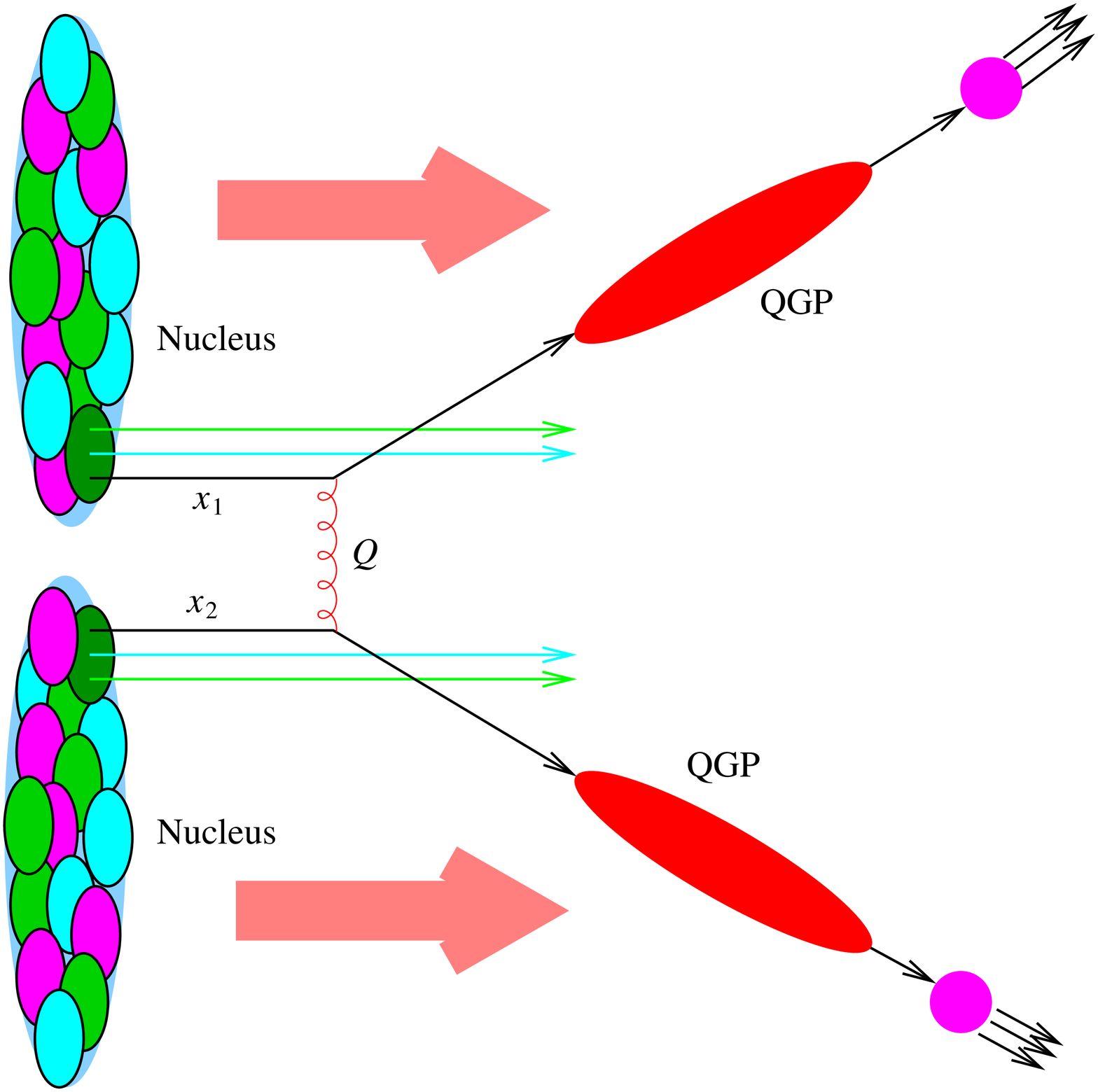}	       

\mbox{} \hfill (a) \hfill (b) \hfill \mbox{}
\caption[]{
Schematic views of hard processes in hadron-hadron collisions (a) and
nucleus-nucleus collisions where QGP is expected to be created (b).
}
\label{fig:schematics}
\end{figure}

The idea behind the jet-quenching phenomenon is rather intuitive. 
As depicted in Figure \ref{fig:schematics}, partons with high $p_T$
(jets) are produced when two
hard partons from colliding hadrons undergo a hard scattering. 
In hadron-hadron collisions (Figure \ref{fig:schematics}-a), the 
jets then propagate and evolve in vacuum until they produce showers of
particles in the detector. On the other hand, in heavy ion collisions 
jets produced by relatively rare hard collisions must propagate and evolve
within the
hot and dense medium (QGP) that is created 
by the rest of the system (Figure \ref{fig:schematics}-b).  
By measuring the difference between the outcome of these two experiments, 
one can then learn about the properties of the medium.  
The simplest consequence of having a dense medium is the 
energy-loss of the propagating parton, that is, ``jet-quenching''. 

The usual way of showing the effect of energy-loss
is to show the ratio of the momentum spectrum from $pp$ scatterings and 
suitably normalized momentum spectrum from $AA$ scatterings.
This ratio is usually referred to as the ``$R_{AA}$'' and defined as
\begin{equation}
R^h_{AA}(p_T)
= 
{(dN^h_{AA}/dp_T)\over N_{\rm binary} (dN^h_{pp}/dpT)}
\end{equation}
where $h$ denotes the observed hadron species and $N_{\rm binary}$ is the
average number of hard binary collisions at the specified impact parameter.
(As far as the present author can find out, the first use of $R_{AA}$ 
appeared in 1990 in publications by M.~Gyulassy 
and M.~Pl\"umer \cite{Gyulassy:1990ye}
and M.~Gyulassy and X.-N.~Wang \cite{Wang:1990bka}.)
Experimental data from PHENIX and STAR collaborations spectacularly confirm
that jet quenching is real. There is a sharp reduction of high energy
particles in the away-side of the trigger particle \cite{Adams:2003im}
and $R_{AA}$ has a surprisingly small value of about $0.2$
for all measured range of the high $p_T$ in central collisions
\cite{Adare:2008qa}.

Theoretically, the calculations of the $pp$ spectrum and the $AA$ spectrum
proceed as follows. For the $pp$ case, the jet cross-section is given by
the following schematic perturbative-QCD formula
\begin{equation}
{d\sigma_{pp}^h\over dt}
=
\int_{abcd}
f_{a/p}(x_a, Q_f)
f_{b/p}(x_b,Q_f)\,
{d\sigma_{ab\to cd}\over dt}\,
D_{h/c}(z_c, Q)
\label{eq:sigmapp}
\end{equation}
where $f_{a/p}$ is the parton distribution function, $d\sigma_{ab\to cd}/dt$
is the parton level cross-section and $D_{h/c}$ is the fragmentation function.
For $AA$ collisions, we have
\begin{equation}
{d\sigma_{AA}^h\over dt} =
{\int_{\rm geometry}}
\int_{abcd}
f_{a/A}(x_a, Q_f) f_{b/A}(x_b,Q_f)\,
{d\sigma_{ab\to cd}\over dt}\,
{{\cal P}(x_c \to x_c'| T, u^\mu)}
D_{h/c'}(z_{c}', Q)
\label{eq:sigmaAA}
\end{equation}
where $f_{a/A}$ is the parton distribution function of the nucleus $A$ and the
conditional probability ${{\cal P}(x_c \to x_c'| T, u^\mu)}$
contains the effect of the medium that changes the parton 
from $c$ to $c'$.
Here $T$ and $u^{\mu}$ denote the 
temperature and the flow velocity profiles throughout the evolution.
We also need to integrate over the nuclear
and collision geometry. Since the geometry is more or less fixed once the
impact parameter range is known, the main task of a theoretician is then 
to calculate
${{\cal P}(x_c \to x_c'| T, u^\mu)}$
given the profiles of the temperature and the flow velocity.

There are, however, some caveats that go with Eq.(\ref{eq:sigmaAA}).
The $pp$ formula (\ref{eq:sigmapp}) is 
firmly based on the factorization theorem in QCD. In contrast, the
factorization theorem that would put 
Eq.(\ref{eq:sigmaAA}) on a firm setting is not yet fully proven.
(Gelis, Lappi and Venugopalan
have taken promising initial steps in this direction 
\cite{Gelis:2008ad,Gelis:2008rw}.)

Another caveat is that the medium is finite and it is evolving all the time.
The lifetime of the QGP created in a heavy ion collision is about 5\,fm/c. 
The size of the system is about 10\,fm. The hydrodynamic time scale is about
$\tau_0 \approx 0.5\,{\rm fm}$. The mean free path of a particle in
a QGP is also of order 1\,fm. None of these are very large or very small 
compared to the others. Full accounting of these similar yet 
different length
and time scales is therefore not an easy task. Inevitably,
one needs to make some assumptions and approximations.

With these caveats, nearly every theoretical approach 
to the energy-loss assumes that the above formula
(\ref{eq:sigmaAA}) is at least a good approximation. 
And that's what will be assumed in this paper as well.

\section{Different Approaches to Parton Energy Loss}

As mentioned above, the main task of a theoretician working on parton
energy loss is to calculate the in-medium modification function 
${{\cal P}(x_c \to x_c'| T, u^\mu)}$.
Collisions with the thermal particles cause the changes 
in the propagating parton.
Hence, the fundamental quantity to calculate is the scattering
cross-sections and the associated collision rates.
If perturbation theory is valid, this would be a relatively straightforward
calculation at least at the leading order.
In a hot and dense matter, this is no-longer true: 
The dispersion relationship of a particle is no longer that in the vacuum. 
The scattering potentials are screened. Extra divergences appear due to the
frequent soft exchanges. All these conspire to make the loop expansion
invalid.

For elastic scatterings, one can still regard the tree diagrams as the
leading order \cite{Mrowczynski:1991da,Mustafa:2003vh}.
For radiational processes, the above complications make
non-trivial resummations necessary even for the leading order.
One must include {\em both} the elastic and the
inelastic scattering processes \cite{Qin:2007rn,Schenke:2009ik}
for phenomenology.
Here, for brevity, we only discuss the inelastic radiational process.

\begin{figure}[ht]
\centering
\includegraphics[width=0.4\textwidth]{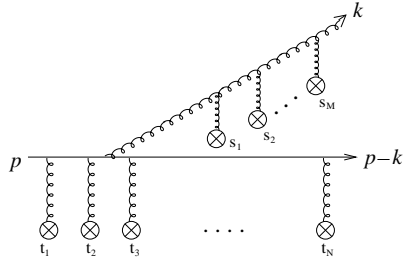}	       

\caption[]{
Schematic views of hard scattering processes in hadron-hadron collisions
(a) and nucleus-nucleus collisions where QGP is expected to be created.
}
\label{fig:the_big_picture2}
\end{figure}

The first study of radiational energy loss was conducted in Refs
\cite{Baier:1996kr,Zakharov:1997uu,Baier:1998kq}.
This method is often referred to as the BDMPS-Z approach.
The main thesis of this approach is as follows.
Consider a medium where the temperature is high enough that perturbation
theory is valid.  In this case, soft exchanges between the medium and the
propagating parton result in the radiation of a hard collinear gluon.
At the same time, the effect of multiple collisions is reduced because
within the coherence length (or formation time) of the emitted gluon, 
all soft scatterings basically count as a single one. This
Landau-Pomeranchuck-Migdal (LPM)
effect then necessitates the resummation of all diagrams depicted in
in Figure \ref{fig:the_big_picture2} in calculating the leading
order gluon radiation rate. 
In the original BDMPS-Z approach 
some simplifications and approximations
were made. Specifically, the medium was assumed to be composed of
random static scatterers. 
Also, the thermal screening effect was only approximately taken into
account using the approach advocated in Ref.\cite{Gyulassy:1993hr}.
The analysis was then carried out in the deep LPM regime and multiple
emission is treated with the Poisson ansatz \cite{Baier:2001yt}.

Subsequent approaches to the energy loss 
calculations use more or less the same starting point.
One still needs to calculate
and sum the diagrams depicted in Figure \ref{fig:the_big_picture2}.
The difference between various approaches may be
classified by the way each approach (i) treats the scattering centers,
(ii) resums the diagrams, and (iii) deals with the evolving medium.
These differences may be summarized as follows. The acronyms in the
following list are the initials of the main authors except HT (Higher-Twist)
and McGill-AMY.
Also, the references given are not exhaustive but just a starting
point for further reading.
\begin{itemize}
\item Scattering centers
 \begin{itemize} 
  \item Heavy static scattering centers (Cold medium) -- 
BDMPS
\cite{Baier:1996kr}, 
Zakharov \cite{Zakharov:1997uu},
GLV, 
\cite{Gyulassy:2003mc,Gyulassy:1999zd}
ASW
\cite{Armesto:2003jh},

  \item Dynamic scatterers (Hot medium) -- 
McGill-AMY \cite{Turbide:2005fk,Qin:2007rn},
DGLV
\cite{Djordjevic:2003zk,Djordjevic:2007at},
WHDG
\cite{Wicks:2005gt}

  \item General nuclear medium with a short correlation length --
HT
\cite{Majumder:2008jy,Majumder:2006we}
 \end{itemize}

\item Resummation schemes
 \begin{itemize}
  \item Sum over diagrams with all possible soft interactions -- BDMPS,
  McGill-AMY
  \item Path integral representation of hard parton propagation -- ASW,
  Zakharov
  \item Reaction operator method with opacity expansion -- GLV, DGLV, WHDG

 \end{itemize}
\item Evolution schemes
 \begin{itemize}
  \item Poisson ansatz -- BDMPS, ASW, GLV, DGLV, WHDG
  \item Fokker-Planck equation -- McGill-AMY
  \item Modified DGLAP equation (momentum space evolution) -- HT 
 \end{itemize}
\end{itemize}
In this proceeding, I am going to use the
McGill-AMY approach as the illustrative example to show what is involved in
calculating each of the above items, simply because that is the
one I am most familiar with.

\section{McGill-AMY Approach}

\begin{figure}[ht]
\centering
\includegraphics[width=0.47\textwidth]{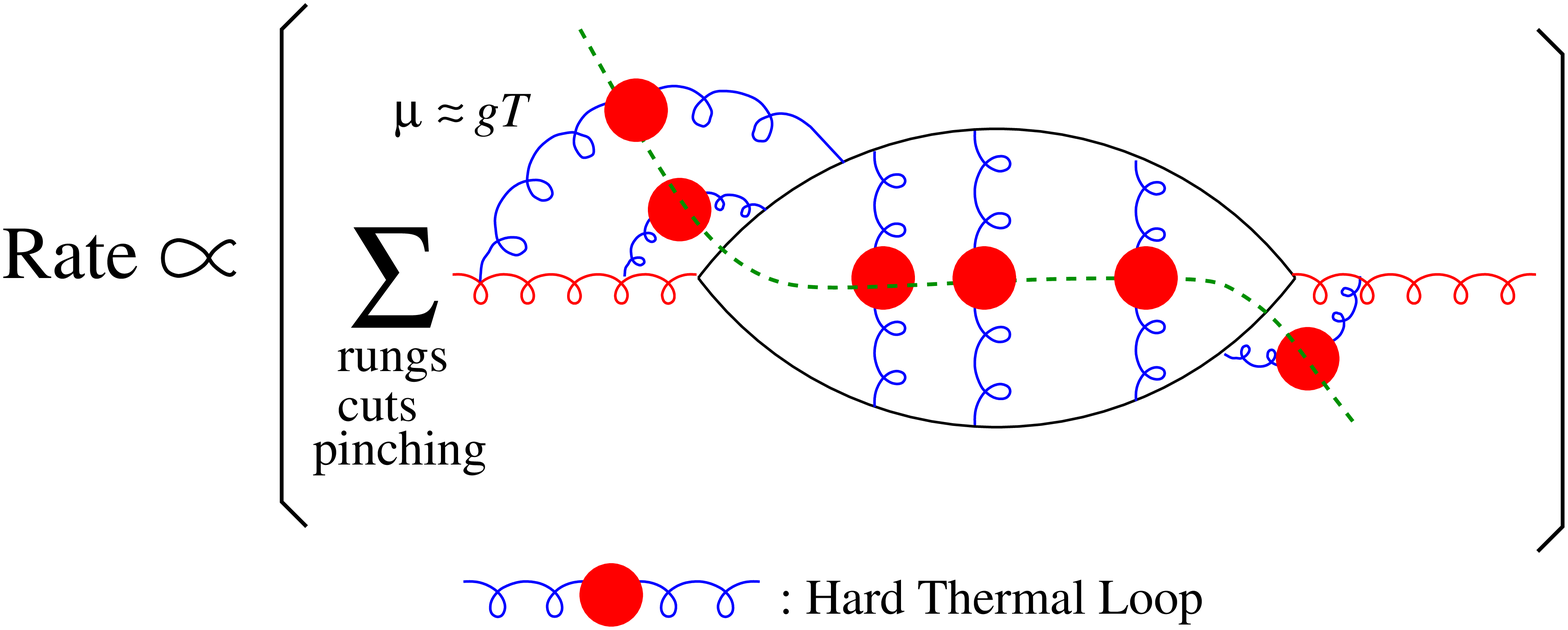}	       
\,\,\,\,\,\,
\includegraphics[width=0.40\textwidth]{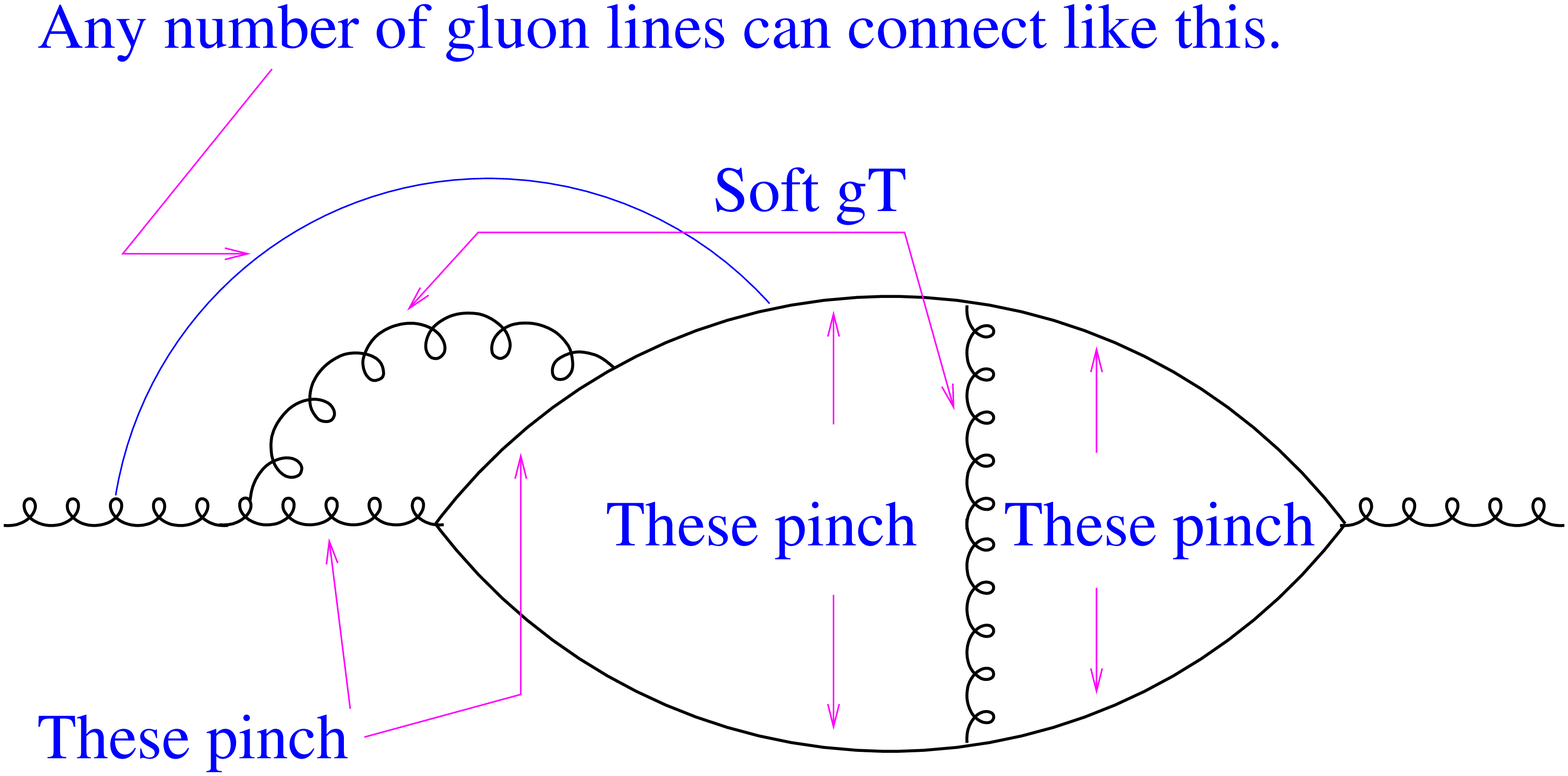}	       

\mbox{} \hfill (a) \hfill (b) \hfill \mbox{}
\caption[]{
Diagrammatic representation of the generalized ladder diagram resummation.
}
\label{fig:rate_calc_amy}
\end{figure}

When the temperature is high enough, 
the asymptotic freedom property of the QCD makes it possible to 
treat QGP within the perturbation theory.
The usual loop expansion, however, is no longer valid. 
The coherence effect (the LPM effect) makes it necessary to resum an
infinite number of generalized ladder diagrams shown in 
Figure \ref{fig:rate_calc_amy}-a. 
Technically, this comes about because connecting any of the three hard lines
with a soft space-like gluon line (labeled with $gT$)
introduces a pair of pinching poles in the loop-frequency space as
illustrated in Figure \ref{fig:rate_calc_amy}-b. The
resulting pinching-pole singularity,
when regulated by the hard-thermal-loop self-energies
cancels the factors of the coupling constants introduced
by the additional interaction vertices. Within thermal field theory, 
Arnold, Moore and Yaffe
rigorously proved that resummation of these generalized ladder diagrams is
ncessary to get the leading order result
(see \cite{Arnold:2002ja} and references therein).
\begin{figure}[ht]
\centering
\includegraphics[width=0.6\textwidth]{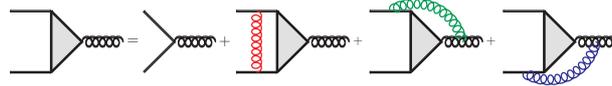}	       
\caption[]{
Diagrammatic representation of the Schwinger-Dyson equation for the radiation
vertices. Figure courtesy of G.~Qin.
}
\label{fig:linear_integral_equation_gluon}
\end{figure}
Luckily, the resummation of the leading order contribution organizes itself
into the Schwinger-Dyson type equation for the radiation vertex. 
Figure \ref{fig:linear_integral_equation_gluon} shows a diagrammatic
representation of the SD equation;
The corresponding integral equation is
\begin{eqnarray}
&&2{\bf h}  = 
i \delta E({\bf h},p,k) {\bf F}({\bf h})
+ g^2 \int \frac{d^2 {\bf q}_\perp}{(2\pi)^2}
C({\bf q}_\perp) 
\times 
\Big\{ 
(C_s-C_{\rm A}/2)[{\bf F}({\bf h})-{\bf F}({\bf h}{-}k\,{\bf q}_\perp)] 
\nonumber\\ &&
\qquad
        + (C_{\rm A}/2)[{\bf F}({\bf h})-{\bf F}({\bf h}{+}p\,{\bf q}_\perp)] 
        +(C_{\rm A}/2)[{\bf F}({\bf h})-{\bf F}({\bf h}{-}(p{-}k)\,
	{\bf q}_\perp)] \Big\} 
\end{eqnarray}
where
\begin{eqnarray}
\delta E({\bf h},p,k) & =&
        \frac{{\bf h}^2}{2pk(p{-}k)} + \frac{m_k^2}{2k} +
        \frac{m_{p{-}k}^2}{2(p{-}k)} - \frac{m_p^2}{2p} \, .
\end{eqnarray}
Here $p$ is the original hard parton momentum and $k$ is the momentum of the
radiated hard gluon.
The masses appearing in the above equation
are the medium induced thermal masses.
The 2-D vector $\bf h$ is defined to be
${\bf h} \equiv ({\bf p} \times {\bf k}) \times {\bf e}_{||}$
where ${\bf e}_{||}$ is the chosen longitudinal direction
and is a measure of the acollinearity.
The differential cross-section 
\begin{equation}
C({\bf q}) = {m_{\rm Debye}^2 \over {\bf q}^2 ({\bf q}^2 + m_{\rm Debye}^2)}
\end{equation}
is the result of the hard-thermal-loop resummation.
The open end of the vertex is then closed off with the bare vertex
and the appropriate statistical factors to get the gluon radiation
rate $\frac{d\Gamma_g(p,k)}{dk dt}$.

In this approach, the medium consists of fully dynamic thermal quarks and
gluons. Furthermore, at least within the perturbation theory, the radiation
rate so calculated is fully leading order in thermal QCD which sums an
infinite number of the generalized ladder diagrams. 

\begin{figure}[ht]
\centering
\includegraphics[width=0.47\textwidth]{raa_radiative_vs_collisional.eps}
\,\,\,\,
\includegraphics[width=0.5\textwidth]{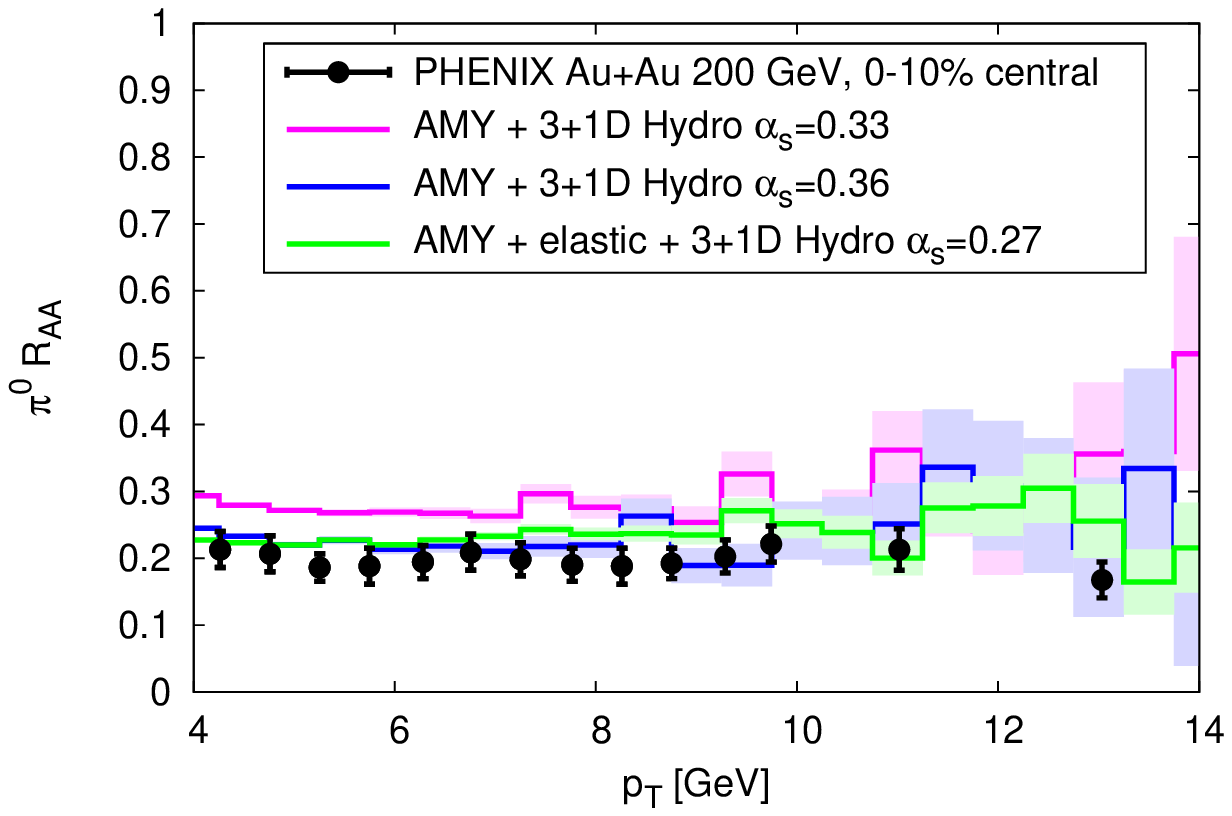}	       

\mbox{} \hfill (a) \hfill (b) \hfill \mbox{}
\caption[]{
The nuclear modification factor $R_{AA}$ for $\pi^0$ calculated within the
McGill-AMY approach.
The panel (a) is from Ref.\cite{Qin:2007rn} and the panel (b) is a
preliminary result from MARTINI -- an event generator being developed based
on the McGill-AMY energy loss mechanism.
In panel (a), 
the red solid line is the full calculation. 
The green broken line includes the effect of radiative energy loss only while
the blue dash-dot line includes the effect of the collisional energy loss
only. 
In panel (b), the green line is the main result with both the collisional
and the radiational energy loss.
Here we set $\alpha_s = 0.27$ and the underlying soft-evolution is
taken from 3+1 D hydrodynamics calculation of Nonaka and Bass 
\cite{Nonaka:2006yn}. The data points are taken from \cite{Adare:2008qa}.
}
\label{fig:raa}
\end{figure}

The original AMY calculation of the radiation rates 
was carried out in order to obtain the leading order transport coefficients 
in hot QGP within the kinetic theory. 
In Ref.\cite{Jeon:2003gi} this
is generalized to the case of propagating hard parton and its kinetic
theory equation.  The initial momentum distributions of
the hard partons now evolve according to the following set of
Fokker-Planck equations
\begin{eqnarray}
\frac{dP_{q\bar{q}} (p)}{dt} \!\!\! &=& \!\!\! \int_k
        P_{q\bar{q}} (p{+}k) \frac{d\Gamma^q_{\!qg}(p{+}k,k)}{dkdt}
        -P_{q\bar{q}} (p)\frac{d\Gamma^q_{\!qg}(p,k)}{dkdt} 
%        \nonumber \\ && \quad
        +2 P_{g} (p{+}k)\frac{d\Gamma^g_{\!q \bar q}(p{+}k,k)}{dkdt}
        \, , \nonumber \\ 
\frac{dP_{g} (p)}{dt} \!\!\! &=& \!\!\!
\!\int_k \!\! 
        P_{q\bar{q}} (p{+}k) \frac{d\Gamma^q_{\!qg}(p{+}k,p)}{dkdt}
        {+}P_{g} (p{+}k)\frac{d\Gamma^g_{\!\!gg}(p{+}k,k)}{dkdt}
      %  \nonumber \\ && \; 
        -P_{g} (p) \left(\frac{d\Gamma^g_{\!q \bar q}(p,k)}{dkdt}
        + \frac{d\Gamma^g_{\!\!gg}(p,k)}{dkdt} \Theta(k{-}p/2) \!\!\right) ,
\nonumber\\
\label{eq:fokker-planck}
\end{eqnarray}
which also includes the effect of absorbing thermal energy from the medium.
For the full phenomenological study, this has to be supplemented by the
local temperature and the flow information by
independent hydrodynamics calculations. Since there is nothing intrinsically
boost-invariant about our formulation, there is no restriction on what the
underlying soft matter evolution should be.
The rates appearing above then become time and space dependent through the
local temperature and the flow velocity.
Finally,
the resulting parton distribution at the final time is convoluted 
with the geometry and the vacuum fragmentation function to yield the medium
modified fragmentation function \cite{Turbide:2005fk,Qin:2007rn}:
\begin{equation}
\tilde{D}(z,Q)
=
\int d^2 s\, {T_A({\bf s})T_B({\bf s}{+}{\bf b})\over T_{AB}({\bf b})}\,
\int dp_f\frac{z'}{z} \left(
{{\cal P}_{\!qq}(p_f;p_i)} D_{q}(z',Q) +
{{\cal P}_{\!g}(p_f;p_i)} D_{g}(z',Q) \right) \, ,
\end{equation}
The resulting $R_{AA}$ for $\pi^0$ is shown in Figure \ref{fig:raa}-a
which includes both the effects of the radiational energy loss and the
collisional energy loss \cite{Qin:2007rn}.
One thing to notice is that $\alpha_s \approx 0.3$ does not validate
perturbative treatment since this implies $g\approx 2$. At this point, 
McGill-AMY approach becomes a phenomenological model.
However, I believe that this is the best one can do 
with current analytical tools
since it at least includes all dynamic
effects such as the energy-momentum conservation,
broadening, thermal push, flavor change, elastic and
inelastic energy losses except the interference between the vacuum and the
in-medium processes \cite{Bass:2008rv}.
The last one is currently
being worked on by the members of the McGill-AMY team.

\section{Discussions}

In the previous section, a particular approach -- McGill-AMY -- was
described and shown to yield results that compare very favorably with the
experimental data. As far as the nuclear modification factor $R_{AA}$ is
concerned, other approaches can equally well describe the experimental
$R_{AA}$ with a reasonable set of assumptions \cite{Bass:2008rv}. 
In this sense, the theoretical efforts to understand the jet quenching has
been successful.
However, the question remains:
What do we really learn about the medium
when different 
approaches with different assumptions about the medium can describe the
experimental data equally well?

For instance, consider the temperature. Since we are dealing with a
thermalized system, finding the value of the temperature goes a long way to
characterize that system. It sounds like measuring temperature ought be a
simple task. Even this, however, becomes a non-trivial task
among different models. Temperature appears as an explicit parameter of the
calculation only in some models.
In others, the medium is treated as dense, but cold to simplify the
calculation. The connection to the QGP property is then made through
the transport coefficient $\hat{q} = \mu^2/\lambda$ where $\mu$ is the
soft momentum exchange scale and $\lambda$ is the mean free path.
When soft exchange dominates the dynamics of the propagating hard parton,
this should not be a bad approximation. 
However, this is still an indirect connection since $\mu$ and $\lambda$ are
independent free parameters instead of functions of $\alpha_s$ and $T$.
Combined with the differing treatments of the multiple emissions,
it makes the determination of the temperature within such models
rather fragile especially since the value of $R_{AA}$ seems to be 
only weakly dependent on the value of $\hat{q}$ \cite{Eskola:2004cr}.

Recently, a collective effort to resolve this and other issues 
has been initiated by the TEC-HQM collaboration \cite{techqm}.
Right now the collaboration is performing
standardized tests to see where exactly the differences between 
models lie. Since the shape of $R_{AA}$ is so featureless, 
I believe that the true test of models will come later when 
other observables such as the hard photon spectrum are calculated 
coherently within each approaches (for instance the $\gamma$ spectrum
\cite{Qin:2009bk,Vitev:2008vk}). 
It is therefore quite encouraging that Monte-Carlo event generators based
on the current works on the energy-loss mechanisms start to appear on the
scene. In these proceedings, some results from 
YaJEM \cite{Renk:2008pp}, 
JEWEL \cite{Zapp:2008gi}, and
Q-PYTHIA \cite{Armesto:2009fj}
are reported. In the remaining space, let me
introduce the event generator being worked on by McGill-AMY team -- MARTINI
(Modular Algorithm for Relativistic Treatment of heavy IoN Interactions).
The first, very preliminary result from MARTINI is shown in Figure
\ref{fig:raa}-b.
MARTINI is a modular modification of PYTHIA 8.1 to take into account the
energy loss of hard partons before they can hadronize (fragment) into
showers. Hopefully, many more tests of the McGill-AMY approach can be
performed on many different hadronic and electro-magnetic probes with this
new development.

%%%%%%%%%%%%%%%%%%%%%%%%%%%%%%%%%%%%%%%%%%%%%
%% end of main text

\section*{Acknowledgments} % 
% please check/modify, comment out or delete if not needed

First of all,
I would like to express my gratitude to the members of McGill-AMY team
-- C.~Gale, G.~Moore, S.~Turbide, G.~Qin, J.~Ruppert, B.~Schenke.
Much help from U.~Heinz, E.~Frodermann, C.~Nonaka, S.~Bass,
M.~Mustafa, and D.~Srivastava is also greatly appreciated.

 % do not change 
\end{document}